\newcommand{\roughly}[1]%
    {{\mathrel{\raise.3ex\hbox{$#1$\kern-.75em\lower1ex\hbox{$\sim$}}}}}

\documentclass[12pt]{article}
\usepackage{paper2e}
\usepackage{epsfig}
\usepackage{latexsym}

\begin{document}
\begin{titlepage}

\preprint{IPMU-09-0099}

\title{Can ghost condensate decrease entropy?}

\author{Shinji Mukohyama}

\address{
Institute for the Physics and Mathematics of the Universe (IPMU)\\ 
The University of Tokyo\\
5-1-5 Kashiwanoha, Kashiwa, Chiba 277-8582, Japan
}

\begin{abstract}

 By looking for possible violation of the generalized second law, we
 might be able to find regions in the space of theories and states that
 do not allow holographic dual descriptions. We revisit three proposals
 for violation of the generalized second law in the simplest Higgs phase
 of gravity called ghost condensate. Two of them, (i) analogue of
 Penrose process and (ii) semiclassical heat flow, are based on Lorentz
 breaking effects, by which particles of different species can have
 different limits of speed. We show that processes in both (i) and (ii)
 are always slower than accretion of ghost condensate and cannot
 decrease the total entropy before the accretion increases the
 entropy. The other proposal is to use (iii) negative energy carried by
 excitations of ghost condensate. We prove an averaged null energy
 condition, which we conjecture prevents the proposal (iii) from
 violating the generalized second law in a coarse-grained sense. 

\end{abstract}

\end{titlepage}

\section{Introduction}

The second law of thermodynamics defines the arrow of time, stating that
entropy does not decrease. Similarly, the attractive nature of gravity
defines the arrow of time for a black hole at least classically. Nothing
can escape from a black hole and, as a result, the area of a black hole
horizon cannot decrease~\cite{Hawking:1971tu}. This classical-mechanical
statement is known as the area law or the second law of black hole
thermodynamics, and it is believed that a black hole has entropy
proportional to the horizon area~\cite{Bekenstein:1973ur}. Quantum
mechanically, however, a black hole emits Hawking
radiation~\cite{Hawking:1974sw} and, hence, the area of a black hole
horizon can decrease. Therefore, the second law of black hole
thermodynamics does not hold quantum mechanically. Instead, it is
believed that the total entropy, i.e. the sum of black hole entropy and
matter entropy outside black hole, does not
decrease~\cite{Bekenstein:1974ax}. This statement about a system
including a black hole and matter is called the generalized second
law. One must, however, be aware that the generalized second law has
been proven only in some limited situations~\cite{Wall:2009wm}.

As already stated, a black hole is believed to have entropy proportional 
to the horizon area. The proportionality coefficient can be determined
by substituting the Hawking temperature for black hole temperature in
the first law of black hole thermodynamics, a relation analogous to the
first law of thermodynamics. The black hole entropy $S_{bh}$ is then
determined as 
%
\begin{equation}
 S_{bh} = \frac{k_Bc^3}{4G_N\hbar}A_h,
\end{equation}
where $A_h$ is the horizon area, $k_B$ is the Boltzmann constant,
$\hbar$ is the Planck constant, $G_N$ is the Newton constant, and $c$ is
the speed of light. The fact that this formula includes $G_N$, $\hbar$
and $k_B$ hints some deep relations among gravity, quantum mechanics and 
statistical mechanics. For this reason, many people believe that black
hole entropy is a key concept towards our understanding of quantum
gravity.

The AdS/CFT correspondence~\cite{Maldacena:1997re}, being one of the
most outstanding recent triumphs of string theory, stemmed from research
in microscopic counting of black hole entropy. It is a concrete
realization of the so called holographic principle that insists
equivalence between a gravitational theory in $d+1$ dimensions and a
non-gravitational field theory in $d$ dimensions. While the AdS/CFT
correspondence applies to gravity with a negative cosmological constant,
it is not known whether there exists a 
holographic principle applicable to gravitational theories with a zero
or positive cosmological constant. For example, it is thought that the
so called dS/CFT correspondence~\cite{Strominger:2001pn}, if it really
exists, would lead to a non-unitary CFT. Moreover, it is believed that a
de Sitter spacetime in string theory is only meta-stable and should
decay into more stable configurations after a certain
timescale~\cite{Kachru:2003aw} (see also \cite{Goheer:2002vf}). How this
could be understood in a field theory dual to de Sitter gravity is not
clear. Since our universe today is thought to have a positive
cosmological constant, it is obviously important to investigate whether
there really exists a holographic principle applicable to gravitational
theories with positive cosmological constant~\footnote{See 
\cite{Sekino:2009kv} for an attempt towards a holographic description of
a theory with exactly zero cosmological constant in background with a
negative spatial curvature.}. Note, however, that absence of holographic
dual would not necessarily imply inconsistencies of a theory or/and a
state.

As a first step towards this outstanding problem, it is intriguing to
try to find a way to identify regions in the space of theories and
states that do not allow holographic dual descriptions. One possible 
strategy is to use the generalized second law. In theories and states
that have holographic dual descriptions, a black hole is presumably dual
to a thermal excitation and, thus, the generalized second law is
expected to be dual to the ordinary second law of
thermodynamics. Therefore, violation of the generalized second law would
indicate lack of holographic descriptions since the ordinary second law
of thermodynamics should hold in non-gravitational theories. For this
reason, by looking for possible violation of the generalized second law,
we might be able to find regions in the space of theories and states
that do not allow holographic dual descriptions.

This approach could be particularly useful for theories which cannot be
embedded in asymptotically anti de Sitter (AdS) spacetime, where the AdS/CFT
correspondence is well formulated. If a gravitational theory is
formulated in asymptotically AdS spacetime then we can analyze and
possibly constrain the theory by using properties of the CFT which is
expected to be dual to it. However, if a theory cannot be embedded in
asymptotically AdS spacetime, then we cannot use this strategy and
should look for other ways. As explained in the previous paragraph, the
generalized second law may provide such a possibility.

Actually, it is known that the simplest Higgs phase of gravity called
ghost
condensate~\cite{ArkaniHamed:2003uy,ArkaniHamed:2005gu,Mukohyama:2006mm}
cannot be embedded in asymptotically AdS spacetime. The reason is very
simple: the coefficient of canonical time kinetic term for excitations
around ghost condensate vanishes if the condensate is spacelike. On the
other hand, in Minkowski and de Sitter backgrounds, ghost condensate is
timelike and gives a healthy time kinetic term to excitations around the 
condensate. It should also be noted that, because of Jeans-like infrared
instability, ghost condensate is not eternal unless cosmological
constant is positive. For this reason, ghost condensate presumably
provides a good testing ground for our strategy using the generalized
second law as a criterion for existence/non-existence of holographic
dual.

In the literature there are already three proposals to violate the
generalized second law by ghost condensate. The purpose of this paper
is to revisit those proposals to see if they really violate the
generalized second law. The conclusion is, unfortunately from the
viewpoint of developing the strategy explained above, that they do not.

The rest of this paper is organized as follows. Sec.~\ref{sec:blackhole}
explains a black hole solution in ghost
condensate. Sec.~\ref{sec:penrose} revisits the proposal by Eling,
Foster, Jacobson and Wall~\cite{Eling:2007qd} based on a classical
process analogous to Penrose process and shows that it does not
decrease the total entropy. In Sec.~\ref{sec:semiclassical} we review
the analysis in \cite{Mukohyama:2009rk}, showing that semiclassical 
heat flow proposed by Dubovsky and Sibiryakov~\cite{Dubovsky:2006vk}
does not violate the generalized second law. In
Sec.~\ref{sec:negativeenergy} we prove a spatially averaged version of
the null energy condition, which we conjecture prevents negative energy
from violating the generalized second law in a coarse-grained
sense. Sec.~\ref{sec:summary} is devoted to a summary of this paper.

\section{Black hole and ghost condensate}
\label{sec:blackhole}

The Schwarzschild metric with the horizon radius $r_g$ is written in the
Lema\^{i}tre reference frame as 
%
\begin{equation}
 g_{\mu\nu}dx^{\mu}dx^{\nu}
  = -d\tau^2 + \frac{r_gdR^2}{r(\tau,R)} + r^2(\tau,R)d\Omega^2,
  \quad r(\tau,R) = \left[\frac{3}{2}\sqrt{r_g}(R-\tau)\right]^{2/3}. 
  \label{eqn:Schwarzschild}
\end{equation}
The vector $\xi^{\mu}$ defined by 
%
\begin{equation}
 \xi^{\mu} = \left(\frac{\partial}{\partial\tau}\right)^{\mu}
  + \left(\frac{\partial}{\partial R}\right)^{\mu}
  \label{eqn:Killingvector}
\end{equation}
satisfies the Killing equation, ${\cal L}_{\xi}g_{\mu\nu}=0$, and is
normalized as 
%
\begin{equation}
 g_{\mu\nu}\xi^{\mu}\xi^{\nu} = - f(r), 
  \quad f(r) \equiv 1-\frac{r_g}{r}. 
\end{equation}

A particle following a radial geodesic is characterized by the mass $m$
and the conserved energy $E$ associated with the Killing vector
$\xi^{\mu}$ as 
%
\begin{eqnarray}
 m^2 & = & -g^{\mu\nu}p_{\mu}p_{\nu}
  = (p_{\tau})^2 - \frac{r}{r_g}(p_R)^2,  \nonumber\\
 E & = & -\xi^{\mu}p_{\mu} = -p_{\tau}-p_R,
\end{eqnarray}
where $p_{\mu}$ is the $4$-momentum covector of the particle. These
equations have two branches of solutions: 
%
\begin{eqnarray}
 p_{\tau} & = & -\frac{E}{f}
  \left[ 1 \pm \sqrt{1-f}\sqrt{1-f\frac{m^2}{E^2}}\right],
  \nonumber\\
 p_R & = & -E - p_{\tau}. 
\end{eqnarray}
For $E>0$, the ``$+$'' sign corresponds to out-going geodesics and
the ``$-$'' sign corresponds to in-coming geodesics. For $E<0$, the
``$-$'' sign corresponds to out-going geodesics and the ``$+$'' sign
corresponds to in-coming geodesics.

Ghost condensate in the Schwarzschild background~\cite{Mukohyama:2005rw}
is approximated by 
%
\begin{equation}
 \phi = M^2 \tau, \label{eqn:phiM2tau}
\end{equation}
where $\phi$ is the scalar field responsible for ghost condensate and
$M$ plays the role of the order parameter of spontaneous Lorentz
breaking. To be precise, the Schwarzschild metric is just an approximate
solution valid within a certain time scale. Actually, ghost condensate
slowly accretes towards the black hole and, as a result, the black hole
mass evolves as~\cite{Mukohyama:2005rw} 
%
\begin{equation}
 M_{bh} \simeq M_{bh0}\times
  \left[ 1 + \frac{9\alpha M^2}{4M_{Pl}^2}
   \left(\frac{3M_{Pl}^2v}{4M_{bh0}}\right)^{2/3} \right],
  \label{eqn:accretion}
\end{equation}
where $M_{bh}$ is the Misner-Sharp energy evaluated at the apparent
horizon, $v$ is the advanced null time coordinate normalized at infinity
(the ingoing Eddington-Finkelstein-type null coordinate), $M_{bh0}$ is
the initial value of $M_{bh}$ at $v=0$, and $\alpha$ ($=O(1)>0$) is a 
coefficient of a higher derivative term. Note that the positivity of
$\alpha$ stems from stability of spatially inhomogeneous excitations of
ghost condensate and, thus, must be respected. The same formula (as well
as the positivity of $\alpha$) applies to the gauged ghost
condensate~\cite{Cheng:2006us}.

\section{Analogue of Penrose process}
\label{sec:penrose}

Eling, Foster, Jacobson and Wall (EFJW)~\cite{Eling:2007qd} proposed a
process analogous to Penrose process to violate the generalized second
law in theories with Lorentz violation. (See also \cite{Kant:2009pm}.)
By applying the EFJW process to a black hole in ghost condensate, one
might think that the generalized second law would be violated. On the
contrary, in this section we shall show that the EFJW process is
inefficient and always dominated over by accretion of ghost condensate.

EFJW consider two classes A and B of particles. Particles in the class A
follow geodesics of the metric $g_{A\mu\nu}$, and those in the class B
follow geodesics of the metric $g_{B\mu\nu}$, where $g_{A,B\mu\nu}$ are
defined up to constant conformal factors as
%
\begin{equation}
 g_{A,B\mu\nu} = -u_{\mu}u_{\nu}
  + c_{A,B}^{-2}(g_{\mu\nu}+u_{\mu}u_{\nu}), 
\end{equation}
$c_{A,B}$ ($c_A\ne c_B$) are positive constants representing limits of
speed, and $u_{\mu}$ is a unit timelike vector representing the
preferred time direction. Without loss of generality, we can assume that 
%
\begin{equation}
 c_A < c_B. 
\end{equation}
Note that, in the following discussions,
ambiguities due to undetermined constant conformal factors can be
absorbed into normalization of mass and energy of particles in each
class. Moreover, only dimensionless quantities such as $E/m$ and $f$ are
important in the following discussions and such ambiguities will cancel
with each other in any physical statements.

In the case of ghost condensate, the preferred direction is specified as
$u_{\mu}=\partial_{\mu}\phi/\sqrt{X}$, where $\phi$ is the scalar field
responsible for the ghost condensate and
$X=-\partial^{\mu}\phi\partial_{\mu}\phi$. Here, it is assumed that
$\partial_{\mu}\phi$ is non-vanishing and timelike. The order parameter
of the spontaneous Lorentz breaking is $M$ defined by the vacuum
expectation value of $X$ as
%
\begin{equation}
 \langle X \rangle = M^4. \label{eqn:XM4}
\end{equation}
For example, see (\ref{eqn:phiM2tau}). Any Lorentz breaking effects,
such as deviation of $c_{A,B}$ from unity, are induced by non-vanishing
$M$ and should vanish in the limit $M^2/M_{Pl}^2\to 0$. Therefore, we
have~\footnote{Since $M_{Pl}$ is defined by $G_N=M_{Pl}^{-2}$, odd
powers of $M_{Pl}$ do not show up. Odd powers of $M$ do not show up
either since $M$ is defined by 
$\langle\partial_{\mu}\phi\rangle\propto M^2$.} 
%
\begin{equation}
 c_{A,B} = 1 + O\left(\frac{M^2}{M_{Pl}^2}\right). 
  \label{eqn:deltacABsmall}
\end{equation}
This is in accord with the fact that quantum corrections via
gravitational interactions generate direct couplings~\footnote{Note that
$\phi$ appears only via its derivatives because of the shift symmetry.}
of matter fields to $\partial_{\mu}\phi$ unless protected by symmetry
and induce Lorentz breaking effects. Since the strength of gravitational
interaction is $G_N=M_{Pl}^{-2}$ and the background value of
$\partial_{\mu}\phi$ is proportional to $M^2$, such induced effects must
be proportional to some positive powers of $M^2/M_{Pl}^2$ at leading
order. The constant conformal factors and the normalization of mass and
energy, mentioned in the previous paragraph, are also
$1+O(M^2/M_{Pl}^2)$.

We suppose that $g_{\mu\nu}$ is the Schwarzschild metric shown in
(\ref{eqn:Schwarzschild}) and that $u_{\mu}=\partial_{\mu}\tau$ (see
(\ref{eqn:phiM2tau}).). In this case, both $g_{A\mu\nu}$ and
$g_{B\mu\nu}$ are Schwarzschild metric with different horizon radii: 
%
\begin{equation}
 g_{A,B\mu\nu}dx^{\mu}dx^{\nu}
  = -d\tau^2 + \frac{r_{gA,B}dR^2}{r_{A,B}(\tau,R)}
  + r_{A,B}^2(\tau,R)d\Omega^2,
\end{equation}
where
%
\begin{eqnarray}
  r_{A,B}(\tau,R) & = & c_{A,B}^{-1}r(\tau,R) = 
  \left[\frac{3}{2}\sqrt{r_{gA,B}}(R-\tau)\right]^{2/3},
  \nonumber\\
  r_{gA,B} & = & c_{A,B}^{-3}r_g.
\end{eqnarray}
The vector $\xi^{\mu}$ defined by (\ref{eqn:Killingvector}) still
satisfies the Killing equation for $g_{A,B\mu\nu}$, 
${\cal L}_{\xi}g_{A,B\mu\nu}=0$, and is normalized as 
%
\begin{equation}
 g_{A,B\mu\nu}\xi^{\mu}\xi^{\nu} = - f_{A,B}(r), 
  \quad f_{A,B}(r) \equiv 1-\frac{r_{gA,B}}{r_{A,B}}
  = 1-c_{A,B}^{-2}\frac{r_g}{r}. 
\end{equation}

Let us prepare two in-coming massive particles outside the horizons: one
in the class A with mass $m_A$ and the other in the class B with mass
$m_B$. Since they are prepared outside the horizon, they have positive
energies, $E_A>0$ and $E_B>0$. We suppose that these particles meet at
$r=r_0$ satisfying $f_A(r_0)<0<f_B(r_0)$,
i.e. in the region between the horizon for the class A and that for the
class B, and split into two massless particles: one in the class A
in-coming with negative energy $E_{A'}<0$ and the other in the class B
out-going with positive energy $E_{B'}>0$. Note that a particle in the
class A can have negative energy at $r=r_0$ since it is inside the
horizon for this class. On the other hand, a particle in the class B
cannot have negative energy at $r=r_0$ since it is outside the horizon
for this class. Since the total energy conserves and $E_{A'}$ is
negative, $E_{B'}=E_A+E_B-E_{A'}$ is larger than the initial total
energy $E_A+E_B$. This is the EFJW process.

Actually, the EFJW process is kinematically forbidden unless
$E_A/m_A=O(\sqrt{f_B})\ll 1$. To see this, note that EFJW invokes
the conservation of momentum covector, which is summarized as
%
\begin{eqnarray}
 E_A + E_B & = & E_{A'} + E_{B'}, \nonumber\\
 p_A + p_B & = &  p_{A'} + p_{B'}. \label{eqn:momentum-conservation}
\end{eqnarray}
Here, 
%
\begin{eqnarray}
 p_{A,B} & = & 
 -\frac{E_{A,B}}{f_{A,B}}
  \left[ 1 - \sqrt{1-f_{A,B}}
   \sqrt{1-f_{A,B}\frac{m_{A,B}^2}{E_{A,B}^2}}\right],
  \nonumber\\
 p_{A',B'} & = & 
  -\frac{E_{A',B'}}{f_{A,B}}
  \left[ 1 + \sqrt{1-f_{A,B}} \right],
\end{eqnarray}
and $f_{A,B}\equiv f_{A,B}(r_0)$. Note that (\ref{eqn:deltacABsmall})
and $f_A<0<f_B$ imply that 
%
\begin{equation}
 f_{A,B} = O\left(\frac{M^2}{M_{Pl}^2}\right). 
  \label{eqn:fABsmall}
\end{equation}
It is easy to solve equations (\ref{eqn:momentum-conservation}) with
respect to $E_{A'}$ as 
%
\begin{equation}
 E_{A'} = \frac{C_AE_A+C_BE_B}{C}, \label{eqn:solEA'}
\end{equation}
where
%
\begin{eqnarray}
 C_A & = & 
  |f_A| + f_B +  |f_A|\sqrt{1-f_B}
  -f_B\sqrt{1+|f_A|}\sqrt{1+|f_A|\frac{m_A^2}{E_A^2}},
  \nonumber\\
 C_B & = & |f_A|\sqrt{1-f_B}
  \left[ 1 + \sqrt{1-f_B\frac{m_B^2}{E_B^2}}\right], \nonumber\\
 C & = &   |f_A|\left[1+\sqrt{1-f_B}\right] 
  + f_B\left[1+\sqrt{1+|f_A|}\right] .
\end{eqnarray}
Here, we have re-expressed $f_A$ as $-|f_A|$. EFJW suppose that 
$E_{A,B}>0$ and $E_{A'}<0$. This indeed requires that $C_A<0$, since
$C_B$ and $C$ are positive definite. This necessary condition is
rewritten as 
%
\begin{equation}
 \frac{E_A^2}{f_B m_A^2} < 
  \frac{1+|f_A|}{2(1+\sqrt{1-f_B})-f_B+\frac{|f_A|}{f_B}(1+\sqrt{1-f_B})^2}
  < O(1),
  \label{eqn:EAsmall}
\end{equation}
and shows that the EFJW process is kinematically forbidden unless 
$E_A/m_A=O(\sqrt{f_B})\ll 1$. This also shows that the initial massive
particle in the class A must be released from a point very close to the
horizon where $r=r_g\times[1+O(\sqrt{f_B})]$. Therefore, before starting
the process, we need to keep such a particle at rest in the vicinity of
the horizon.

It is also easy to see that (\ref{eqn:solEA'}) combined with $E_A>0$ and
$E_B>0$ implies that 
%
\begin{equation}
 -\frac{E_{A'}}{\sqrt{f_B}m_A}
  < \frac{\sqrt{f_B}\sqrt{1+|f_A|}}{C}
  \sqrt{\frac{E_A^2}{m_A^2}+|f_A|}. 
\end{equation}
This, combined with (\ref{eqn:EAsmall}), leads to 
%
\begin{equation}
 -\frac{E_{A'}}{\sqrt{f_B}m_A} < O(1). \label{eqn:EA'small}
\end{equation}
This means that the amount of negative energy carried by the final
massless particle in the class A is rather low: 
$|E_{A'}|/m_A=O(\sqrt{f_B})\ll 1$.

EFJW treats all particles participating the process as test
particles. Therefore, in order to justify this treatment their
gravitational backreaction to the geometry must be small enough. Let
$\Delta r\simeq (|f_A|+f_B)\ r_g$ be the difference between horizon
radii for the class A and the class B. The corresponding proper distance
is $\Delta l\simeq 2\sqrt{r_g\Delta r}$. In order to justify the test
particle treatment, $\Delta l$ must be sufficiently longer than the
gravitational radius of each massive particles. This requires that 
%
\begin{equation}
 m_{A,B} \ll \frac{1}{2}M_{Pl}^2 \Delta l 
  \simeq 2M_{bh}\sqrt{|f_A|+f_B}. 
  \label{eqn:mABsmall}
\end{equation}
As already stated at the end of the paragraph before the last, before
starting the process, we need to keep the initial particle with mass
$m_A$ at rest at $r=r_g\times[1+O(\sqrt{f_B})]$. By demanding that this 
initial condition should not disturb the geometry in the vicinity of the
horizon, we obtain a similar condition on $m_A$. 
%
\begin{equation}
 m_A \ll M_{bh}\times O(\sqrt{f_B}). 
  \label{eqn:mAsmall}
\end{equation}

Using (\ref{eqn:fABsmall}) in (\ref{eqn:EA'small}) and
(\ref{eqn:mABsmall}) (or (\ref{eqn:mAsmall})), we obtain 
%
\begin{equation}
 |E_{A'}| \ll M_{bh}\times O\left(\frac{M^2}{M_{Pl}^2}\right). 
  \label{eqn:EA'toosmall}
\end{equation}
This shows that the amount of negative energy sent to the black hole by
the EFJW process is rather low. As EFJW states, it takes time of order
$r_g$ to perform this process since particles need to travel this
distance~\footnote{As already stated, the initial massive particle in
the class A should be kept at rest near the horizon. We need to supply
it from somewhere else and somehow stop it near the horizon. This
already takes time scale of order $r_g$.}. During this time scale, ghost 
condensate accretes into the black hole, according to the
formula~(\ref{eqn:accretion}). This amounts to the increase of the black
hole mass given by 
%
\begin{equation}
 \Delta M_{bh}|_{v=r_g} \sim 
  M_{bh}\times \frac{M^2}{M_{Pl}^2}. 
\end{equation}
This is much larger than the amount of negative energy
(\ref{eqn:EA'toosmall}). Therefore, we conclude that the EFJW process is
too inefficient to decrease black hole entropy in ghost condensate. This
conclusion trivially extends to gauged ghost condensate since the
accretion rate is the same.

\section{Semiclassical heat flow}
\label{sec:semiclassical}

For the Schwarzschild background (\ref{eqn:Schwarzschild}), the effective
metric $g_{A\mu\nu}$ for particles in the class A and the effective
metric $g_{B\mu\nu}$ for particles in the class B have different surface
gravity and, thus, different Hawking temperatures $T_{bhA}$ and
$T_{bhB}$. Without loss of generality, we can assume that
$T_{bhA}<T_{bhB}$.

By using the semiclassical heat flow due to Hawking radiation, Dubovsky
and Sibiryakov (DS)~\cite{Dubovsky:2006vk} proposed a process to violate
the generalized second law in ghost condensate. DS consider two shells
surrounding the black hole, one with temperature $T_{shellA}$
interacting with particles in the class A only and the other with
temperature $T_{shellB}$ interacting with particles in the class B
only. By tuning these temperatures of the shells, one can satisfy 
%
\begin{equation}
 T_{bhA} < T_{shellA} < T_{shellB} < T_{bhB},
  \label{eqn:temperatures}
\end{equation}
and ensure that the net energy flux from the shell A to the black hole
is equal to the net energy flux from the black hole to the shell B. In
this case, energy is transfered from the shall A to the shell B via the
black hole while black hole mass remains unchanged. Since the shell A
has lower temperature than the shell B, this process appears to violate
the generalized second law. This is the DS process.

As shown in \cite{Mukohyama:2009rk}, however, the DS process is
suppressed by the factor $M^2/M_{Pl}^2$ and, as a result, is much slower
than the Jeans instability of ghost condensate. Indeed, it is even
slower than accretion of ghost condensate~\footnote{In
\cite{Mukohyama:2009rk}, comparison with the time scale of accretion was
explicitly illustrated for gauged ghost condensate only, but it holds
also for ungauged ghost condensate since the quoted accretion rate is
common for gauged and ungauged ghost condensate.}. Therefore, black hole 
entropy increases due to accretion before the DS process starts
operating. Here, we shall briefly review the argument of
\cite{Mukohyama:2009rk}.

As already stated, the scale $M$ is the order parameter of spontaneous 
Lorentz breaking and the Lorentz symmetry should recover in the 
$M^2/M_{Pl}^2\to 0$ limit. This is the reason why the deviation of
limits of speed from unity is suppressed by $M^2/M_{Pl}^2$, as shown in
(\ref{eqn:deltacABsmall}). This implies that differences among various 
temperatures in (\ref{eqn:temperatures}) are suppressed by
$M^2/M_{Pl}^2$. In particular, we have 
%
\begin{equation}
 \left|T_{shellA,B}-T_{bhA,B}\right|
  = T_{bh}\times O\left(\frac{M^2}{M_{Pl}^2}\right), 
\end{equation}
where $T_{bh}$ is the temperature of the metric $g_{\mu\nu}$, and the
net energy flux from the shell A or B to the black hole is 
%
\begin{equation}
 F_{shell\to bh} \sim \pm T_{bh}^2 \times
  O\left(\frac{M^2}{M_{Pl}^2}\right), 
  \label{eqn:netfluxsmall}
\end{equation}
where the ``$+$'' sign is for the shell A and the ``$-$'' sign is for
the shell B. Note that the net energy flux from each shell to the black
hole vanishes when temperatures of the black hole and the shell agree,
i.e. in the limit $M^2/M_{Pl}^2\to 0$.

DS argue that the sum of entropy of the shell A and entropy of the shell
B decreases since energy moves from the shell A with lower temperature
to the shell B with higher temperature. However, the temperature
difference is again suppressed by $M^2/M_{Pl}^2$ and, thus, the rate of
decrease of shells' entropy is 
%
\begin{equation}
 \frac{dS_{shells}}{dt} 
  = \left(\frac{1}{T_{shellB}}-\frac{1}{T_{shellA}}\right)
  \left|F_{shell\to bh}\right|
  \sim -\frac{\left|F_{shell\to bh}\right|}{T_{bh}}\times 
  O\left(\frac{M^2}{M_{Pl}^2}\right). 
\end{equation}
Combining this with (\ref{eqn:netfluxsmall}), we obtain
%
\begin{equation}
 \frac{dS_{shells}}{dt} 
  \sim -T_{bh}\times 
  O\left(\frac{M^4}{M_{Pl}^4}\right). 
\end{equation}
This is highly suppressed. Indeed, it takes the time scale $t_{DS}$
defined by 
%
\begin{equation}
 t_{DS} \sim T_{bh}^{-1}\times \frac{M_{Pl}^4}{M^4}
\end{equation}
for the DS process to decrease shells' entropy just by one unit.

Now, the formula~(\ref{eqn:accretion}) tells us that the black hole 
entropy significantly increases by accretion of ghost condensate in the 
time scale $t_{DS}$. On the other hand, the shells' entropy can decrease
just by one unit in this time scale. Therefore, the total entropy
including the black hole entropy increases and the DS process does not
violate the generalized second law~\cite{Mukohyama:2009rk}. This
conclusion trivially extends to gauged ghost condensate since the
accretion rate is the same.

\section{Negative energy}
\label{sec:negativeenergy}

As a yet another proposal to violate the generalized second law, let us
consider negative energy carried by excitations of ghost
condensate~\cite{ArkaniHamedTalkPI}. In this section we shall see that
an averaged energy condition holds, and we conjecture it protects the
generalized second law.

Let us consider the action of the form
%
\begin{equation}
 I = \int dx^4\sqrt{-g}P(X), \quad
  X= - \partial^{\mu}\phi\partial_{\mu}\phi.
  \label{eqn:action}
\end{equation}
The stress-energy tensor is 
%
\begin{equation}
 T_{\mu\nu} = (\rho+P)u_{\mu}u_{\nu} + Pg_{\mu\nu}, 
\end{equation}
where
%
\begin{equation}
 \rho = 2P'X - P, \quad 
  u_{\mu} = \frac{\partial_{\mu}\phi}{\sqrt{X}}. 
\end{equation}
The equation of motion is 
%
\begin{equation}
 \nabla^{\mu}J_{\mu} = 0, 
  \label{eqn:eom}
\end{equation}
where 
%
\begin{equation}
  J_{\mu} \equiv -2P'\partial_{\mu}\phi
\end{equation}
is the current associated with the shift symmetry and the corresponding
charge is conserved.

Ghost condensate is characterized by a non-vanishing timelike vacuum
expectation value of $\partial_{\mu}\phi$ as in (\ref{eqn:XM4}). In the
language of the action (\ref{eqn:action}), it corresponds to the value
of $X$ where $P'=0$. Actually, $P'=0$ is a dynamical attractor of the
system in an expanding universe. In the flat Friedmann-Robertson-Walker
(FRW) background the equation of motion (\ref{eqn:eom}) for a
homogeneous $\phi$ is $\partial_t(a^3P'\partial_t\phi)=0$ and implies
that 
%
\begin{equation}
 P'\partial_t\phi \propto \frac{1}{a^3}\to 0 \quad (a\to\infty),
\end{equation}
where $a$ is the scale factor of the universe. There are two choices:
$P'\to 0$ or $\partial_t\phi\to 0$. The later corresponds to the trivial
Lorentz invariant background and the former corresponds to the ghost 
condensate. The ghost condensate is, in this sense, a dynamical
attractor of the system. Fluctuations around the ghost condensate 
background obtain a time kinetic term with the correct sign if 
$P''>0$. On the other hand, the leading spatial kinetic term comes from
higher derivative terms such as $(\Box\phi)^2$. For the correct sign of
the higher derivative term, those fluctuations are stable and have a
healthy low energy effective field theory. Hence, ghost condensate is
characterized by the background $X=M^4$ ($>0$) satisfying $P'(M^4)=0$
and $P''(M^4)>0$. The scale $M$ is the order parameter of spontaneous
Lorentz breaking and also plays the role of the cutoff scale of the low
energy effective field theory for excitations of ghost
condensate~\footnote{Extra modes due to higher time derivative terms
have frequencies around or above $M$ and, thus, are outside the regime
of validity of the low energy effective field theory. }.

In the ghost condensate background, $\rho+P=2P'X$ vanishes and thus
$T_{\mu\nu}$ acts as a cosmological constant. This is the reason 
why Minkowski and de Sitter spacetimes are exact solutions in ghost
condensate~\footnote{Anti de Sitter spacetime is not a solution in ghost
condensate with higher derivative terms, essentially because there is no
flat FRW slicing in anti de Sitter spacetime. Spacelike condensate leads
to instability of excitations around the condensate.}. If we consider
fluctuations around the ghost condensate background then we notice that 
$\rho+P=2P'X\simeq 2M^4P''(M^4)\delta X+O(\delta X^2)$, where 
$\delta X=X-M^4$. This means that $\rho+P$ is positive for $\delta X>0$ 
and negative for $\delta X<0$. Therefore, the null energy condition can
be violated by excitations of ghost
condensate.

In the following, however, we shall prove a spatially averaged version
of the null energy condition.

The Lagrangian $P$ is expanded as
%
\begin{equation}
 P = M^4\left[p_0 + \frac{1}{2}p_2\chi^2 + O(\chi^3)\right], \quad
  \chi \equiv \frac{X}{M^4}-1,
\end{equation}
where $p_0=P(M^4)/M^4$ and $p_2=M^4P''(M^4)=O(1)>0$. Thus, we obtain 
%
\begin{equation}
 \rho+P - M^2 J_{\mu}u^{\mu}
  = 2P'X \left( 1 - \frac{M^2}{\sqrt{X}}\right)
  = M^4\left[p_2\chi^2 + O(\chi^3)\right]. 
  \label{eqn:rhoPJu}
\end{equation}
In the regime of validity of the effective field theory, $|\chi|\ll 1$
and the right hand side is non-negative. Therefore, by integrating
(\ref{eqn:rhoPJu}) over a spacelike hypersurface orthogonal to
$u^{\mu}$, we obtain 
%
\begin{equation}
 \int d\Sigma (\rho+P) \geq M^2 Q, 
\end{equation}
where $Q$ is the conserved charge associated with the shift symmetry: 
%
\begin{equation}
 Q = \int d\Sigma J_{\mu}u^{\mu}. 
\end{equation}

As stated in the third paragraph of this section, $P'=0$ is a dynamical
attractor in an expanding universe. Thus, it is very natural to set
$Q=0$. Moreover, if $Q<0$ today then $P'$ was negative with large
absolute values in the past. Since a large negative $P'$ leads to UV
instabilities of the system, negative $Q$ is strongly  
disfavored. Actually, with the exact shift symmetry, it is impossible to
have a negative $Q$ without introducing UV instabilities~\footnote{If
the shift symmetry is softly broken then it is possible to have a
negative $Q$ without instabilities in expanding
universe~\cite{Creminelli:2006xe}.}. Therefore, it is necessary to
assume that the shift charge $Q$ initially vanishes or starts with a
positive value~\footnote{Negative energy solutions presented in 
\cite{Feldstein:2009qy} are inconsistent with any initial conditions
with $Q\geq 0$ and thus are excluded by this condition.}. In this case
we obtain 
%
\begin{equation}
 \int d\Sigma (\rho+P) \geq 0. 
\end{equation}
This is the averaged null energy condition.

The averaged energy condition states that negative energies are always
accompanied by larger positive energies. This is somehow similar to the
so called quantum inequalities and the quantum interest
conjecture~\cite{Ford:1999qv} in ordinary quantum field theory: even in
Minkowski spacetime the ordinary field theory can have negative local
energy, but negative energy is always accompanied by larger positive
energy. Since direct couplings between ghost condensate and matter 
fields are suppressed by the Planck scale (cf. the third paragraph of
Sec.~\ref{sec:penrose}), excitations of ghost condensate interact with
ordinary matter only gravitationally. This implies that those positive
and negative energies cannot be separated from each other by
hand. Therefore, although one could decrease black hole entropy by 
gravitationally exciting a lump of negative energy and sending it into a
black hole, larger positive energy should follow it and eventually
increase the black hole entropy. For this reason, we conjecture that the
averaged energy condition protects the generalized second law in a
coarse-grained sense~\footnote{In ordinary thermodynamics, entropy can
fluctuate both upwards and downwards by order unity but will eventually
increase. Thus, in ordinary thermodynamics the second law holds only in
a coarse-grained sense.}.

\section{Summary}
\label{sec:summary}

We have revisited three proposals to violate the generalized second law
by ghost condensate: (i) analogue of Penrose process, (ii) semiclassical
heat flow, and (iii) negative energy. The proposals (i) and (ii) are
based on Lorentz breaking effects, by which particles of different
species can have different limits of speed. We have shown that processes 
in both (i) and (ii) are always slower than accretion of ghost
condensate and cannot decrease the total entropy before the accretion 
increases the entropy. We have also proved an averaged null energy
condition, which we conjectured prevents the proposal (iii) from
violating the generalized second law in a coarse-grained sense.

\section*{Acknowledgements}

The author would like to thank Brian Feldstein and Simeon Hellerman for 
useful discussions. The work of the author was supported in part by MEXT
through a Grant-in-Aid for Young Scientists (B) No.~17740134, by JSPS
through a Grant-in-Aid for Creative Scientific Research No.~19GS0219,
and by the Mitsubishi Foundation. This work was supported by World
Premier International Research Center Initiative (WPI Initiative), MEXT,
Japan.


\end{document}